# *AWiMA*: An architecture for *A*dhoc *W*ireless *M*obile internet-*A*ccess

Dilip Krishnaswamy (dilip@ieee.org)
Advanced Technology R&D Group, Office of the Chief Scientist, Qualcomm Inc.
5665 Morehouse Drive, Qualcomm Research Center, QRC-603U, San Diego CA 92121

**ABSTRACT**
This paper suggests a system architecture for wireless wide-area-networking access using adhoc networking between a mobile Client node without direct connectivity to a wireless-wide-area-network and a mobile Service Provider node with connectivity to a wireless-wide-area-network. It provides a means for securely providing such adhoc wireless networking services using a Server for tunneling and routing, registration and authentication. The architecture also provides support for handoff of a Client node from one Service Provider to another with persistence of a tunnel between the Client and the Server enabling a soft-handoff. Different wireless protocols may be used for adhoc networking, with filtered interconnection of authenticated Clients implemented at a Service Provider node. The architecture is applicable across different wide-area-network protocols, and provides simultaneous support for multiple wide-area-network protocols.

## 1 INTRODUCTION

Wireless wide area networks provide broadband Internet access to mobile subscribers over a regional, a nationwide, or even a global region. However, accessing WWANs from all mobile devices may not always be possible. Some mobile devices may not have the WWAN radio needed to access a WWAN at a given location. Other mobile devices with a WWAN radio may not have a subscription plan enabled to access the WWAN. A significant fraction of wireless capacity is going unutilized and every second that goes by where such available capacity is not used is lost opportunity for mobile clients to use the available spectrum to access the internet. It is also lost opportunity for wireless network operators and carriers to provide such service to mobile clients. Adhoc wireless networking allows mobile devices to dynamically connect over wireless interfaces using protocols such as WLAN (Wireless Local Area Network) [2], Bluetooth, UWB (Ultra Wide Band) or other wireless protocols. This paper addresses a pressing need in current mobile systems and networks for the existence of architectures that would allow a user of a mobile device without WWAN access to securely and dynamically subscribe to wireless access service provided by a user with a WWAN-capable mobile device using wireless adhoc networking between the mobile devices belonging to the two users. The paper suggests an architecture to enable such a capability. Trust is provided in the architecture through the use of a common server node. The paper also addresses handoff considerations within the scope of the suggested architecture to enable handoff of client nodes from one service provider node to another service provider node.

## 2 *AWiMA* SYSTEM ARCHITECTURE

Consider a mobile node with internet access using a wireless-wide-area-networking (WWAN) (such as LTE/UMB/WiMAX/UMTS/EVDORevAorB/GRPS/EGPRS/HSDPA/HSUPA/other) technology/protocol [1]. This WWAN-capable mobile node can create in its vicinity an adhoc network based on the same or based on an alternative wireless technology (for example, using WLAN (802.11a/b/g/n/s) or UWB or Bluetooth or other short range wireless technologies). The mobile node has the ability to transport packets between different wireless interfaces, for example, between a WLAN radio interface and a WWAN radio interface. This enables the mobile node to provide internet access for the adhoc clients connected to the mobile node. Adhoc clients can connect to this mobile node over the WLAN radio interface, and can then share the internet connection available over the WWAN radio interface. Such a mobile node providing such a service is called a Service Provider. The goal is to extend the availability of the internet to neighboring nodes that can talk to the Service Provider. A neighboring node that uses the internet access service provided by the Service Provider is called the Client. Given the number of different wireless protocols that are available for the wireless backhaul, it is possible that Client nodes may come into contact with Service Provider nodes with access to different WWAN protocols. The capacity of the wireless link for the backhaul may vary based on the protocol used, and the modulation and coding scheme used over the wireless link. The dynamic wireless conditions over the wireless link for the backhaul can vary significantly as well. Due to the mobility of Service Providers and Clients, handoff of a Client from one Service Provider to another may be required. It is necessary for such handoff to be possible across different wireless technologies for the backhaul. A Client should be able to move from one Service Provider with a given wireless WAN technology for backhaul to a different Service Provider with a different wireless WAN technology for backhaul. This requires a node that is higher up in the architecture relative to the Service Providers and the Clients that can help in managing such handoff. This node is called the Server node. The Server node resides anywhere on the internet such that all wireless WAN infrastructure networks can connect to it. The Server node helps in registering Clients and Service Providers and authenticates them as well. A secure tunnel is established between the Client node and the Server node. The persistence of the tunnel during handoff allows for different WWAN technologies to coexist in the system architecture while allowing for a handoff of a Client node across WWAN technologies supported by the Service Provider nodes.

It should be noted that a Service Provider could use any WWAN protocol or network operator for its internet-access,

and to provide service by extending its wireless connectivity using any adhoc networking protocol. It is expected that revenue generated from the service will be shared between the different entities involved in providing the service, including the Service providers, the Server, and the wireless network operators and carriers involved in providing wireless infrastructure access.

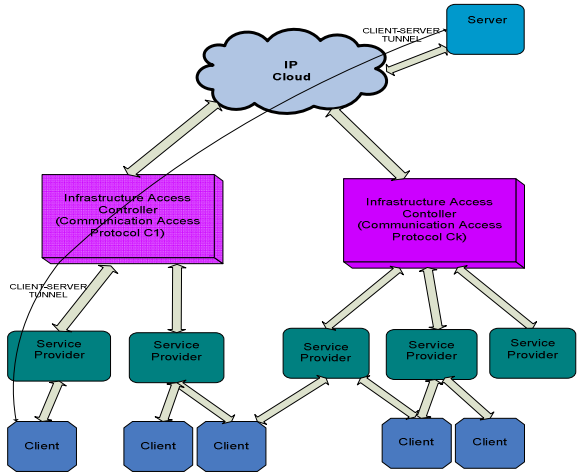

**Figure 1:** *AWiMA* **System Architecture**

Figure 1 depicts the overall *AWiMA* system architecture

## 3 KEY COMPONENTS

The system architecture includes Service Providers, Clients, and a Server.

### 3.1 The Server

The Server can be implemented in a centralized or distributed manner. The Server provides services to register new Clients and Service Providers, to dynamically authenticate Clients and Service Providers, to maintain data tunnels with Clients, and to monitor information regarding sessions, and determine revenue distribution policies. The Server contains information, such as a database for Clients and Service Providers, and information required for AAA/payment information. The Server allows for tunnels to be established between it and the Clients, and provides NAT and routing of packets from/to Clients to/from the internet. Different services (such as tunneling and authentication) at the server can be provided by different geographically-located server nodes. A connectivity graph with geographic information of Clients and Service Providers can be created at the Server based on feedback from Clients and Service Providers to the Server regarding the beacon information of Service Provider nodes in their vicinity.

Based on the needs (duration/performance) of a Client and the availability/link conditions/mobility/energy-level of a Service Provider, the Server can optionally provide support for handoff of a Client from one Service Provider to another Service Provider.

A Service Provider quality metric is stored about the quality of service provided by each Service Provider; this information can be provided to Clients who may want to choose from available Servers. This metric is continuously updated as more information becomes available about a specific Service Provider node. If $G_{prev}$ is the previous value of the goodness metric, $G_{session}$ is the goodness value for the current session hosted by a Service Provider, then the new goodness metric $G_{new}$ can be computed using an equation such as $G_{new} = \alpha G_{session} + (1 - \alpha) G_{prev}$, where $0 \leq \alpha \leq 1$. The Server also determines how to allocate revenue generated from Clients to all the elements in the value chain (such as the Service Provider, and the wireless network operator/carrier providing service to the Service Provider).

#### 3.1.1 The Data Tunneling Anchor

The location of the data tunneling anchor for the client data tunnel session could be at the server. Alternatively, it could also be located closer to the client or somewhere within the WWAN infrastructure network of the wireless network operator/carrier, whereas the authentication could be performed external to the WWAN infrastructure network.

### 3.2 The Service Provider

The Service Provider extends its wireless broadband internet-access service to neighboring nodes which are charged for using this extended service. Network address translation is required to convert the private addresses in the wireless local area network to be changed to the public IP address of the Service Provider for outbound traffic, and for the reverse mapping to be performed for inbound traffic. When the Service Provider decides to provide such service, it sends a request to the Server for approval to provide such service. After authentication and approval by the Server, the Service Provider can advertise an adhoc WLAN SSID (Service Set Identifier) to allow other adhoc nodes (potential Clients) with WLAN radios to connect to it. Service Providers are allowed to create multiple adhoc networks based on different wireless protocols such as WLAN or UWB or Bluetooth or other technologies. Clients can choose to select which adhoc network to use to connect to the Service Provider. Figure 2 shows a typical Service Provider node with filtered interconnection of traffic related to authenticated adhoc Clients between the adhoc wireless links to the WWAN interface. The Service Provider admits Clients and provides them with a certain QoS guarantee (such as an expected average bandwidth during the session, and average time of availability of access). Platform intelligence on the Service Provider that projects the increased utilization of available battery based on the additional service provided by would be useful in determining whether should consider providing such service at a given location at a given time based on platform constraints at that time. The Service Provider can include support for manual control for session management with regard to the Client connectivity, and session monitoring. Alternatively, the Service Provider can support a seamless operation mode where a defined fraction of its network/platform resources are provisioned for services such that the services are provided by the Service Provider node without the need for active participation of the user of the Service Provider node. A Service Provider may provision some of its resources for handoff. It can allow a fraction of its resources for lightweight sessions with potential Clients that may be serviced by other Service Providers.

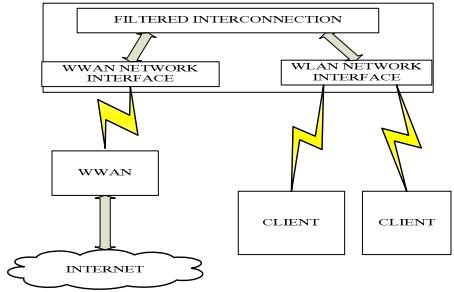

**Figure 2: Service Provider Node**

### 3.2.1 Service Provider Utility Function

In general, a service provider may choose to provide service or choose to handoff a client by computing a utility function based on various parameters such as the amount of energy available on the platform, the duration for which it is available, its desire to improve its goodness metric, the amount of traffic consumed by local applications on the service provider platform, the additional energy consumption rate to provide service, the revenue associated with providing the service, and the utility related to supporting other clients. If the Service Provider dynamically determines that it is unable to provide service for the duration of the service that was agreed to with the Client, then it notifies both the Server and the Client regarding its unavailability. The Server can then consider a handoff of the Client to another Service Provider, if there is such a Service Provider in the vicinity of the Client.

### 3.3 The Client

A Client looks for available Service Providers. When it detects the presence of one or more Service Providers, then it can initiate a connectivity session with a Service Provider based on parameters such as the available bandwidth that the Service Provider can support, the "goodness" metric of the Service Provider, and the cost/duration of the service advertised. The Client can obtain such information from the Service Provider beacons, with static information for the session (such as the goodness metric) in SSID names, and dynamically changing information in vendor-specific fields in a Service Provider's beacon. Alternatively, it can obtain such information by connecting to the Service Provider and obtaining a custom message from the Service Provider. Additionally, a Client can connect through one Service Provider, and request information from the Server about all Service Providers in its vicinity. The Client may have its own MobileIP [3] or IPv6 address. Alternatively, it may be receive a temporary DHCP address or a MobileIP or IPv6 address from the service provider node or the server node. A VPN session is initiated at the Client to connect to the VPN Server at the Server so that all application data between the Client and the Server is encrypted. Handoff is possible with a Client maintaining lightweight connectivity sessions with multiple Service Providers, while using one Service Provider for its internet connectivity service. The Server authenticates potential Service Providers to enable session handoff. The Client does not necessarily have to maintain lightweight sessions with other Service Providers; a Client implementation could consider a handoff only when necessary, and could just be aware of the list of potential alternative Service Providers. When handoff is not possible, a Client has to reconnect via a different Service Provider to reinitiate internet connectivity. Handoff support can be provided from one Service Provider to another across network operators and across wireless infrastructure access protocols.

With the tunnel between the Client and the Server in place, all Client applications see the Client's VPN IP address as the source IP address and the IP address of the internet destination address as the destination IP address. In the IP header for the tunnel, the DHCP address of the Client in the adhoc network is the source IP address, while the IP address of the Server is the destination address. As data flows through the Service Provider, the Client DHCP IP address is NATed to the public IP address of the Service Provider, and port mapping is done to a different port, so that reverse traffic can be correctly mapped to the appropriate Client. Since applications at the Client only use the ClientVPN address, the persistence of the tunnel during handoff enables soft handoff in the architecture. Tunneling could be implemented using a persistent tunneling technology such as OpenVPN [4].

### 3.3.1 Client Node Utility Function

Client nodes may have access to different Service Provider nodes. In such a case, a Client node may evaluate a utility function to select a Service Provider based on parameters such as the cost of the service, duration of available service, goodness metric of the servicer provider, average WWAN backhaul bandwidth available from the service provider, and the wireless link quality between the client and the service provider. The client may also choose a combination of service providers, if available, to meet its session performance requirements. In such a case, a client node utility function needs to account for the possibility of fractional allocation of client data traffic to each service provider node.

### 3.3.2 Parallel Sessions Management

If the bandwidth needs of a Client are greater than the capabilities of each of the Service Providers providing service, then the Client could set up simultaneous connections with multiple Service Providers. A Client with multiple radios could potentially connect to multiple Service Providers simultaneously using a different radio for each connection. It could also connect to the same Service Provider using multiple radios over different wireless channels and/or protocols. If connections are maintained using the same wireless protocol, then different channels or OFDM sub-channels could be used to have parallel session support. If the Client has only one radio available for such connectivity, then it has to distribute the time connected with each Service Provider, which can be accomplished with proper scheduling of transmissions.

### 4 Security Considerations

Security is a key aspect of the overall architecture. Pairwise session keys $K_{SP,S}$, $K_{C,S}$ and $K_{SP,C}$ are set up for *custom control messages* to be exchanged between the Service Provider and the Server, between the Client and the Server, and between the Service Provider and the Server, respectively. All Client data is encrypted and tunneled to/from a Tunneling Anchor via the Service Provider. Standard protocols such as EAP-TLS or

EAP-TTLS may be used to provide security in the system. A security architecture is suggested in the following sub-sections. It should be noted that techniques such as Diffie-Hellman may also be used for session establishment.

### 4.1 Service Provider-Server Control Session

The Service Provider requests a certificate from the Server. Upon receipt of the certificate, and after validating the Server certificate, the Service Provider suggests a session key ($K_{SP,S}$) encrypted with the public key of the Server. This is received by the Server, decrypted with its private key, and all subsequent messages are encrypted with the session key $K_{SP,S}$. The Service Provider provides its credentials encrypted with the session key $K_{SP,S}$ to allow the Server to authenticate it. A secure control session $X_{SP,S}$ is then established between the Service Provider and the Server using the key $K_{SP,S}$.

### 4.2 Client-Server Control Session

Client dynamic authentication requires connectivity over an ad-hoc wireless link between the Client and the Service Provider. The Client requests and receives a certificate from the Server via the Service Provider. After validating the Server certificate, the Client suggests a session key ($K_{C,S}$) encrypted with the public key of the Server. This is received by the Server, and all subsequent messages between the Server and the Client are encrypted with the session key $K_{C,S}$. The Client provides its credentials encrypted with the session key $K_{C,S}$ to allow the Server to authenticate the Client. The Server authenticates the Client and informs the Service Provider and the Client that the Client is now in an authenticated state to receive service. A secure control session $X_{C,S}$ established between the Client and the Server using the key $K_{C,S}$.

### 4.3 Client – Service Provider Secure Control Session

A key $K_{SP,C}$ is generated at the Client. This is communicated to the Server over the session $X_{C,S}$ (communication of data for the session $X_{C,S}$ proceeds through the Service Provider node). The Server then communicates the key $K_{SP,C}$ to the Service Provider over the session $X_{SP,S}$. Now both the Client and the Service Provider have the key $K_{SP,C}$. This establishes a secure control session $X_{SP,C}$ between the Client and the Service Provider using the key $K_{SP,C}$. It is possible that the key $K_{SP,C}$ is generated either at the Server, or at the Service Provider or at the Client.

### 4.4 Client-Server Data Tunnel

Over the secure session $X_{C,S}$, information can be exchanged between the Client and the Server to establish an encrypted VPN tunnel $T_{C,S}$ to transport data to the internet through the tunneling anchor Server. All data from the Client destined to any location on the internet is tunneled through the Server. This is done to ensure that the Service Provider has no visibility into Client data, and hence ensures privacy of the Client. This tunneling also provides protection to the Service Provider by ensuring that all Client data flows through the Server, leaving the responsibility of such Client transactions to the Server and the Client, with the Service Provider merely serving as a transport to allow Client data to reach the Server.

### 4.5 Client-Service Provider Adhoc Wireless Link

To prevent any visibility into lower layer protocol-stack information flowing over the adhoc wireless link between the Client and the Service Provider, the Client and the Service Provider can agree to a wireless link key $WK_{SP,C}$ (for example to provide WPA2 encryption).

### 4.6 Secure Handoff

It is possible that a handoff is desired from one Service Provider SP1 to another Service Provider SP2. Let us assume that the Client is connected through Service Provider SP1 to the Server. This implies the existence of three secure control sessions $X_{SP1,S}$, $X_{C,S}$ and $X_{SP1,C}$ using keys $K_{SP1,S}$, $K_{C,S}$ and $K_{SP1,C}$ respectively. When a Service Provider SP2 becomes available, a secure session $X_{SP2,S}$ gets established using a key $K_{SP2,S}$. A handoff request may be initiated by either the Client or the Server or the Service Provider SP1. Over the secure session $X_{SP2,S}$ the Server can provide information to Service Provider SP2 that the Client is an authenticated one. Over the secure session $X_{C,S}$, the Client is informed by the Server that it has been authenticated with the Service Provider SP2. A key $K_{SP2,C}$ is generated either at the Client or the Server or Service Provider SP2 for the establishment of a secure session $X_{SP2,C}$. The Client disassociates with Service Provider SP1 and associates with Service Provider SP2. Key $K_{SP2,C}$ is used for the secure session $X_{SP2,C}$ between the Client and the Service Provider SP2. Information (such as residual client packets) can be exchanged between the service providers through the common server node for both service providers. Alternatively, such exchange of information can happen over a direct wireless link or multihop wireless path between the service providers, if the service providers can reach each other over a local wireless link/path. A key $K_{SP1,SP2}$ is established for secure exchange of messages between the service providers.

### 4.7 Parallel Session Management

When a client uses multiple service providers simultaneously, multiple control sessions are set up ($X_{C,S}$, $X_{SP1,S}$, $X_{SP1,C}$ $X_{SP2,S}$, and, $X_{SP2,C}$ for two service providers for example). However, there is only one client VPN data tunnel $T_{C,S}$ such that client-related data packets travel between the tunneling anchor and the client through any of the service providers simultaneously supporting the client. Based on the dynamic capabilities and constraints of each service provider, client traffic may be dynamically reallocated across the existing service providers or partially moved to new service providers.

## 5 Handoff Considerations

Handoff in the architecture is depicted in Figure 3. Handoff can be initiated by either the client or the service provider or by the server. A client node may communicate an active list of potential target service providers to the server. The server pre-authenticates (this term is used in the architecture to keep a target service provider ready to provide service to a client with session keys established in advance) these target service providers. Packets that have left the mobile client may be in transit to the current mobile service provider, or could be at the current mobile service provider. These packets need to continue to be supported by the current mobile service provider. Other packets that have left the mobile client may be in transit to the server, or may be waiting at the server for further processing, or may be in transit to their final destination beyond the tunneling server. Future packets that leave the mobile client are sent to the target mobile service provider after handoff. Packets that are destined to the mobile client may be waiting at the server. Such packets are sent to the target mobile service provider after handoff. Other packets destined for the mobile client may be in transit to the current mobile service provider, or may be waiting at the current mobile service provider, or may be in transit from the current service provider to the mobile client, and the current mobile service provider needs to continue to support such packets to be delivered to the mobile client. The delivery of such packets can be done over a wireless link or a multi-hop wireless path between the current mobile service provider and the target mobile service provider. Alternatively, such packets can be delivered by the current mobile service provider to the server, which then sends them through the target mobile service provider. It is possible that some information (such as control flow information) may go through the server node, while other information (such as data flow information) may go over the direct wireless link/path between the service providers.

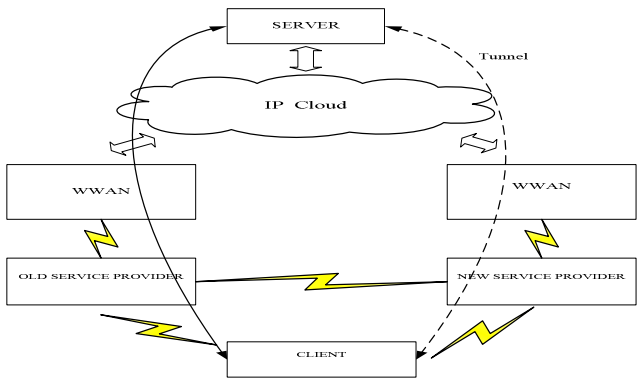

**Figure 3: Handoff Options**

Thus, during handoff, messages between the current mobile service provider and the target mobile service provider may be exchanged either through the server, or over a wireless link or multi-hop wireless path between the service providers. Timers at the current service provider node could be used to determine when to stop providing such a service. It is possible that the current service provider may become completely unavailable; in this case, the end-to-end sessions that survive through the client-server tunnel are responsible for recovering packets if a reliable transport protocol such as TCP is used. For unreliable transport protocols such as UDP or UDP-lite, it is possible that some packets traveling through the older service provider are lost. However, one could consider outer coding such as Reed Solomon coding across groups of packets, so that the information packets can be recovered if there is sufficient redundancy across packets associated with a given flow.

## 6    Conclusion

To conclude, the goal of the *AWiMA* system architecture is to support internet access for mobile clients using adhoc quasi-static/mobile service provider nodes in heterogeneous wireless networks. The architecture enables wireless WAN access using adhoc networking between a mobile Client node without WWAN connectivity and one or more adhoc Service Provider nodes with WWAN connectivity. It provides a means for securely providing such services using a Server for tunneling and routing, registration and authentication. The architecture also provides support for handoff of a Client node from one Service Provider to another. The architecture is applicable across different WWAN protocols, and provides simultaneous support for multiple WWAN protocols. Different wireless protocols may be used for adhoc networking, with filtered interconnection of authenticated Clients implemented at a Service Provider node.

It is also possible that a Service Provider node may need to take on Client functionality depending on the availability of service for the wireless protocol and the network operator's service that the Service Provider may have subscribed to for its WWAN access. The suggested architecture will allow such a Service Provider node (now without WWAN access) to access the internet using the WWAN access provided by another Service Provider node.

Additional options could include multi-hop connectivity where a Client node may extend the service, acting as a secondary Service Provider, to another Client node that cannot connect directly to a primary Service Provider node based on link conditions; in this case, the intermediate Client node would behave both as a Service Provider and as a Client.

From a game-theoretic perspective, dynamic variations in parameters related to utility functions of Service Providers and Clients can lead to interesting non-cooperative games with dynamically varying Nash equilibria at a given geographic location. Future research could explore such equilibria.

Over a period of time, Client-only nodes may transition over and become Service Providers in the system architecture. A dynamic equilibrium may be achieved with a fluid heterogeneous network and with mobile Clients and Service Providers where Clients continue to exist due to a reduced cost of dynamic service provided by the architecture and the occasional needs of the Client.

**ACKNOWLEDGEMENTS**

The author would like to thank Atul Suri and Rob Daley for discussions and for encouraging this research. The author would like to thank Saurabh Oberoi, Vladimir Bychkovsky, and Dhaval Shah for their help with a prototype implementation. Mark Charlebois and Vidya Narayanan are also thanked for discussions.